\documentclass{svjour2}                    
\smartqed  
\usepackage{graphicx}
\usepackage{newtxtext,newtxmath}
\usepackage{bm}
\usepackage{amsmath}
\usepackage{dsfont}

\usepackage{url}
\usepackage[colorlinks=true,breaklinks=true,allcolors=blue]{hyperref}

\usepackage{cite}

\usepackage{geometry}
\newcommand\ZZ{{\mathds{Z}}}

\newcommand{\mvec}[1]{\boldsymbol #1}

\DeclareMathOperator{\Tr}{{Tr}}
\DeclareMathOperator{\erf}{{erf}}

\journalname{Journal of Statistical Physics}
\begin{document}

\title{Subexponentially growing Hilbert space and nonconcentrating distributions in a constrained spin model
}


\author{Jason R. Webster         \and
        Michael Kastner 
}


\institute{Jason R. Webster \at
              Institute of Theoretical Physics, Department of Physics, University of Stellenbosch, Stellenbosch 7600, South Africa
           \and
           Michael Kastner \at
              National Institute for Theoretical Physics (NITheP), Stellenbosch 7600, South Africa, and\\
              Institute of Theoretical Physics, Department of Physics, University of Stellenbosch, Stellenbosch 7600, South Africa\\
              \email{kastner@sun.ac.za}
}

\date{Received: date / Accepted: date}

\maketitle

\begin{abstract}
Motivated by recent experiments with two-component Bose--Einstein condensates, we study fully-connected spin models subject to an additional constraint. The constraint is responsible for the Hilbert space dimension to scale only linearly with the system size. We discuss the unconventional statistical physical and thermodynamic properties of such a system, in particular the absence of concentration of the underlying probability distributions. As a consequence, expectation values are less suitable to characterize such systems, and full distribution functions are required instead. Sharp signatures of phase transitions do not occur in such a setting, but transitions from singly peaked to doubly peaked distribution functions of an ``order parameter'' may be present.
\end{abstract}

\section{Introduction}

Equilibrium statistical mechanics, as originally developed by Boltzmann and Gibbs in a classical context \cite{Boltzmann,Gibbs}, makes predictions about the equilibrium expectation values of observables on the basis of weighted averages over phase space or Hilbert space. In a quantum mechanical $N$-body system with finite-dimensional Hilbert space, the dimension of that space usually grows exponentially with $N$, and this exponential growth can be traced back to the product structure of the underlying $N$-body Hilbert space. The exponential growth is essentially responsible for the logarithm in Boltzmann's famous formula for the entropy, $S=k\log W$, and has many further implications on statistical mechanics. A few theoretical examples where classical  configuration space grows faster than exponentially with $N$ have been discussed in the mathematical literature \cite{Dhar78,Jensen_etal}, but the link to physical models is not evident. To devise a many-body system with a subexponentially growing configuration space or Hilbert space, an obvious strategy is to impose a sufficient number of constraints such that the usual exponential growth is impeded. While this is easily achieved mathematically, it is less obvious how to realize such constraints in physically realistic models or even experimentally realizable systems.

In this paper we discuss an example of a system with a subexponentially growing Hilbert space dimension, realized experimentally by means of a Bose--Einstein condensate. In the setting of the experiments described in Refs.~\cite{Ziebold_etal10,Gross_etal10,Strobel_etal14}, a condensate of \textsuperscript{87}Rb atoms is coherently distributed between two internal states of the ground state manifold. Irradiating the sample with resonant two-photon radio-frequency microwaves generates a linear coupling between the internal states, while $s$-wave scattering between the atoms induces a nonlinear interaction. If both internal states are in the same spatial mode, and after applying a Schwinger mapping to represent the bosonic operators in terms of pseudospins, the dynamics of such a condensate consisting of $N$ bosons can be described by the Hamiltonian
\begin{equation}\label{e:fcTFIM}
H=-\frac{J}{2N}\sum_{i,j=1}^N \sigma_i^x \sigma_j^x - \frac{h}{2}\sum_{i=1}^N \sigma_i^z.
\end{equation}
The real constants $J$ and $h$ are determined by parameters of the atomic transitions as well as of the microwave and laser fields (see \cite{Ziebold_etal10,Gross_etal10} for details). By $\sigma_i^x$ we denote the $x$ component of the Pauli operator acting on site $i$ of a lattice of $N$ sites. The lattice is a fully connected graph, as all pairs of lattice sites interact with each other at equal strength. For that reason it is convenient to introduce collective spin operators
\begin{equation}
S^a=\frac{1}{2}\sum_{i=1}^N\sigma_i^a,\qquad a\in\{x,y,z\},
\end{equation}
which allow us to express the Hamiltonian as
\begin{equation}\label{e:Hcollective}
H=-2J(S^{x})^2-2hS^{z}.
\end{equation}
Hamiltonian \eqref{e:fcTFIM} or, equivalently, \eqref{e:Hcollective} can be seen as a fully-connected version of the transverse-field Ising model (TFIM), and it is also a special case of the exactly solvable Lipkin--Meshkov--Glick (LMG) model \cite{LipkinMeshkovGlick65}, which has been studied in a variety of contexts \cite{Gilmore84,Cirac_etal98,VdovinStorozhenko99,TsayTzeng_etal99,HeissScholtzGeyer05,BarthelDusuelVidal06,RibeiroVidalMosseri07,CanevaFazioSantoro08,Scherer_etal09,Kastner10,KastnerJSTAT10,OlivierKastner14,Campbell_etal15,Opatrny_etal15,Hauke_etal16,SantosTavoraPerezBernal16}.
The operator $\mvec{S}^2 = (S^{x})^2 + (S^{y})^2 + (S^{z})^2$ commutes with the Hamiltonian, and the latter is therefore block-diagonal. A block of size $2S+1$ is associated to each spin quantum number $S$, where $S\leq N/2$. Within each block, we label the different eigenstates and eigenenergies $E_{S,\alpha}$ by the additional quantum number $\alpha$.

In several of the above references on the LMG model, finite-temperature properties are investigated \cite{Gilmore84,VdovinStorozhenko99,TsayTzeng_etal99,Scherer_etal09,Kastner10,KastnerJSTAT10,OlivierKastner14,Hauke_etal16}, based on calculations of, for example, the canonical partition function $Z$ as a function of the inverse temperature $\beta$,
\begin{equation}\label{e:Z_LMG}
Z(\beta)=\sum_{S}^{N/2}\sum_\alpha d_S \exp\left(-\beta E_{S,\alpha}\right),
\end{equation}
where the sum over $S$ starts at $0$ if $N$ is even, and at $1/2$ otherwise. The weight $d_S$ in \eqref{e:Z_LMG} is a combinatorial factor, defined by the multiplicity in which the individual spin-$1/2$ degrees of freedom in \eqref{e:fcTFIM} can be arranged into a collective spin-$S$; see \cite{Scharf72,KastnerJSTAT10} for details. 
Finite-temperature properties based on the partition function \eqref{e:Z_LMG}, like the phase diagram in Fig.~3 of \cite{Hauke_etal16}, have been advocated as being linked to the above described experimental realization of the fully-connected TFIM by means of a two-component Bose--Einstein condensate \cite{Ziebold_etal10,Gross_etal10,Strobel_etal14}. The problem here, however, is the fact that, while such a condensate is indeed described by the Hamiltonian \eqref{e:fcTFIM} or \eqref{e:Hcollective}, its finite-temperature behavior is not described by the partition function \eqref{e:Z_LMG}. This is due to the fact that the condensate is subject to the additional constraint of being in a fully symmetric state or, in the spin language, of being in the maximum spin sector $S=N/2$. This symmetry is a consequence of the fact that the spatial part of the condensate wave function is symmetric, which necessitates the internal degrees of freedom of the two-component gas to be permutation symmetric. Taking this constraint into account, the canonical partition function is
\begin{equation}\label{e:Z_LMGconstrained}
Z(\beta)=\sum_\alpha \exp\left(-\beta E_{N/2,\alpha}\right),
\end{equation}
where we have made use of the fact that $d_{N/2}=1$. As we will discuss in the remainder of this paper, thermodynamic properties of such a constrained spin model differ substantially from those of the unconstrained model. In particular, the Hilbert space dimension grows only subexponentially (in fact, linearly) with the system size $N$, unlike the commonly encountered exponential growth which is responsible for many familiar results in statistical mechanics.

The present paper arose from discussions with experimentalists, wondering whether it is possible to use the \textsuperscript{87}Rb realization of the LMG Hamiltonian to experimentally probe the phase diagram of the LMG model, and in particular its signatures in the quantum Fischer information \cite{Hauke_etal16}. To answer this question, we study in this paper the thermal equilibrium properties of the LMG model subject to the additional constraint $S=N/2$. As we will see, the constraint makes it necessary to resort to an alternative definition of the ``thermodynamic limit.'' Moreover, in this limit, we report unconventional large-$N$ behavior, particularly manifest in the absence of concentration of measure, i.e., the failure of the relevant probability distributions to become exponentially sharp in the large-system limit $N\to\infty$. In this sense, the large-$N$ limit does not even lead to proper thermodynamic behavior, and for this reason we will speak of the {\em large-$N$ limit}\/ in the following (avoiding the term {\em thermodynamic limit}). The absence of concentration of measure has substantial implications, and does not allow for phase transitions in the usual sense to occur. Although no sharp transition between different phases occurs, we are able to identify regimes in parameter space that are distinguished by their qualitative features of distribution functions of observables. Identifying these phases requires access to these distribution function, which is more challenging than identifying conventional phase transitions through, e.g., their order parameter. The resulting ``phase diagram'' of the thus defined transition is not in any way related to the well-known phase diagram of the LMG model without a constraint to the maximum spin sector. 

Before presenting these results in detail, we would like to point out that the \textsuperscript{87}Rb experiments in which a fully-connected TFIM Hamiltonian with the constraint $S=N/2$ is realized \cite{Ziebold_etal10,Gross_etal10,Strobel_etal14} emulate the unitary time evolution of an {\em isolated}  system. As yet, no experimental results are available for the fully-connected TFIM immersed in a thermal bath for which the canonical partition function \eqref{e:Z_LMGconstrained} would be an adequate description. We will comment on the feasibility of such an experimental realization in the concluding Sec.~\ref{s:Conclusions}.

\section{Density of states of the fully-connected XX model}
\label{s:DoS}

We write the Hamiltonian of the LMG model \cite{LipkinMeshkovGlick65} in the form
\begin{equation}\label{e:LMGcollective}
H=-\frac{2}{N}\left[(S^{x})^2+\gamma(S^{y})^2\right]-2hS^{z},
\end{equation}
where the constant $J$ in \eqref{e:Hcollective} has been set to $1$ to fix the energy scale. The $1/N$ prefactor in \eqref{e:LMGcollective} is not always included in the definition, but it is a convenient way of rescaling energies such that the large-$N$ limit of the model is well-defined and yields nontrivial results. The anisotropy parameter $\gamma$ permits interpolation between the fully-connected TFIM \eqref{e:Hcollective} at $\gamma=0$ and the fully-connected XX model at $\gamma=1$, i.e., the case of isotropic interactions in the $xy$-plane. This latter case is particularly amenable to analytic calculations, and for that reason we will focus on it for the larger part of this paper. The fully-connected TFIM, which is realized in the above described experiments with Bose-Einstein condensates, behaves qualitatively very similar, as will be discussed in Sec.~\ref{s:fcTFIM}. For $\gamma=1$, the Hamiltonian of the fully connected XX model can be written as
\begin{equation}\label{e:XXcollective}
H=-\frac{2}{N}\left[\mvec{S}^2-(S^{z})^2\right]-2hS^{z},
\end{equation}
which commutes with both, $\mvec{S}^2$ and $S^{z}$. Hence the angular momentum eigenstates $|S,M\rangle$, satisfying
\begin{align}
\mvec{S}^2|S,M\rangle&=S(S+1)|S,M\rangle,\\
S^{z}|S,M\rangle&=M|S,M\rangle,
\end{align}
form a complete eigenbasis of the Hamiltonian \eqref{e:XXcollective}. Taking into account the constraint $S=N/2$ for the spin quantum number, we can write the Hamiltonian eigenvalue equation as
\begin{equation}
H|N/2,M\rangle=E_{M,h}|N/2,M\rangle
\end{equation}
with
\begin{equation}
E_{M,h}=-\frac{N}{2}-1+\frac{2M^2}{N}-2hM.
\end{equation}

\begin{figure}\centering
  \includegraphics[width=0.55\linewidth]{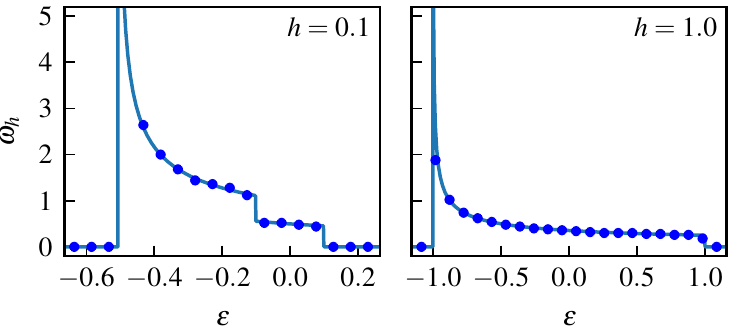}
  \caption{
  Density of states $\omega_h$ as a function of $\varepsilon$ for the fully-connected XX model \eqref{e:XXcollective} of system size $N=500$ and for different values of the magnetic field $h$. The dots mark numerical results, obtained by directly evaluating \eqref{e:Omegah} with $\Delta=0.1$. The continuous lines show the integral approximation \eqref{e:omegaXX}, in excellent agreement with the numerical results for all parameter values studied.}
	\label{f:dosnum}
\end{figure}

The microcanonical density of states of the constrained model is given by
\begin{equation}\label{e:Omegah}
\Omega_h(\varepsilon)=\sum_{M=-N/2}^{N/2}\delta_\Delta(N\varepsilon-E_{M,h}),
\end{equation}
where $\varepsilon$ is the energy per lattice site. The function
\begin{equation}\label{e:deltaDelta}
\delta_\Delta(x)=\begin{cases}
1/\Delta & \text{for $x\in [-\Delta,0]$,}\\
0 & \text{otherwise},
\end{cases}
\end{equation}
(with small but otherwise arbitrary $\Delta>0$) is used in \eqref{e:Omegah} to count the number of states in an energy range $[E,E+\Delta E]$. 
The microcanonical density of states $\Omega_h$ is a convenient starting point for the computation of all kinds of averages and probability distributions of relevance in statistical mechanics, including those of other statistical ensembles (like the canonical or grandcanonical one). It inherits the $N$ dependence from the underlying Hilbert space, and for this reason $\Omega_h$ is expected to grow linearly with $N$. This suggests the definition of
\begin{equation}
\omega_h(\varepsilon)=\frac{1}{N}\Omega_h(\varepsilon),
\end{equation}
which is expected to be convergent in the limit $N\to\infty$. In this limit, and assuming $\Delta$ to be very small, we can calculate the sum in \eqref{e:Omegah} by integral approximation,
\begin{equation}\label{e:omegaXX}
\begin{split}
\omega_h(\varepsilon)&\simeq\int_{-N/2}^{N/2}\delta(N\varepsilon-E_{M,h})dM\\
&= \begin{cases}
2\omega_0(\varepsilon,h) &\text{for $\varepsilon\in [B_0,B_1]$ and $|h|\leq 1$},\\
\omega_0(\varepsilon,h) & \text{for $ \varepsilon\in [B_1,B_2]$},\\
0 & \text{otherwise},
\end{cases}
\end{split}
\end{equation}
where $\delta$ denotes the Dirac distribution and
\begin{equation}
\omega_0(\varepsilon,h)=\frac{1}{2\sqrt{h^2+2\varepsilon+1+2/N}}.
\end{equation}
The boundaries $B_i$ are defined by
\begin{subequations}\begin{align}
B_0&=-\frac{h^2}{2}-\frac{1}{2}-\frac{1}{N},\\
B_1&=-|h|-\frac{1}{N},\\
B_2&=|h|-\frac{1}{N}.
\end{align}
\end{subequations}
In accordance with the $\ZZ_2$ symmetry of the Hamiltonian \eqref{e:XXcollective}, the density of states \eqref{e:omegaXX} is symmetric in $h$, and we can restrict the following discussion to $h\geq0$. As illustrated in Fig.~\ref{f:dosnum}, the result in \eqref{e:omegaXX}, although obtained in integral approximation, agrees remarkably well with exact numerical results, even for moderate system sizes of $N=500$. This convergence of $\omega_h$ in the limit of large $N$ confirms the above conjectured linear (and hence subexponential) growth of the density of states $\Omega_h$ with system size, which implies that the entropy, according to its conventional definition
\begin{equation}\label{e:sublinS}
S_h(N,\varepsilon)\simeq\ln\left[N\omega_h(\varepsilon)\right] =\ln N+\ln\omega_h(\varepsilon),
\end{equation}
is nonextensive. Some people would claim that this is a problem that has to be cured by resorting to a different definition of the entropy, but we contend that through such an {\em ad hoc}\/ modification one loses, or discards, essential physical features of the system under investigation. 

\begin{figure}[b]\centering
  \includegraphics[width=0.95\linewidth]{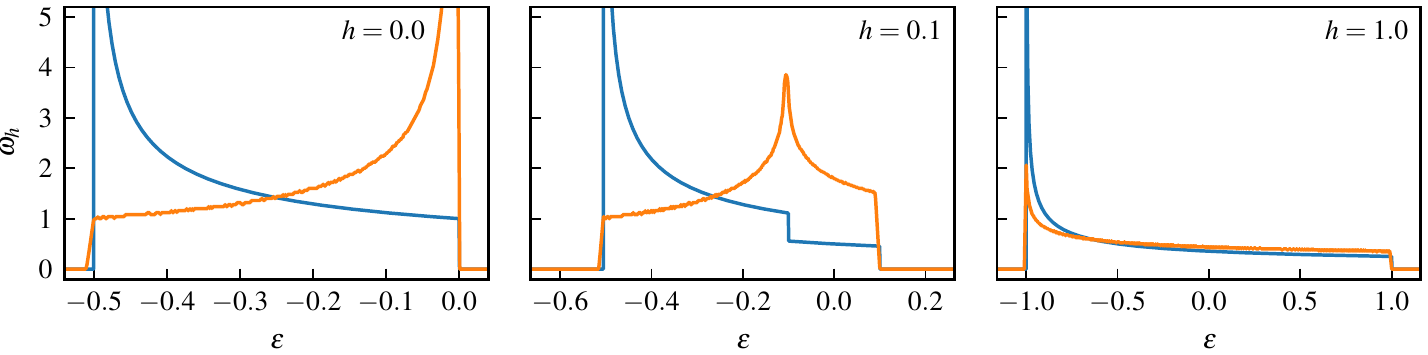}
	\caption{\label{f:TFIM_dos} 
	Density of states $\omega_h$ as a function of the energy density $\varepsilon$. The blue lines show results for the fully-connected XX-model in the continuum approximation \eqref{e:omegaXX}. The orange lines show data for the fully-connected TFIM, calculated according to \eqref{e:Omegah} by making use of the numerically computed eigenvalues of the Hamiltonian \eqref{e:TFIMcollective} for parameter values $N=5000$ and $\Delta=0.01$.
	Unlike for the fully connected XX model, the peak of the fully-connected TFIM, which diverges in the infinite-$N$ limit, is not constrained to the ground state energy, but can move through the spectrum when $h$ is varied.}
\end{figure}

For $|h|\leq1$, the density of states \eqref{e:omegaXX} in the continuum limit shows a divergence at the ground state energy, visible as a sharp peak in all three plots of Fig.~\ref{f:TFIM_dos}. These divergences have been discussed in the literature in the context of so-called excited-state quantum phase transitions \cite{CaprioCejnarIachello08,SantosTavoraPerezBernal16}. In a finite system no such divergence is present, but a sharp finite peak remains. 

\section{Canonical probability distributions of the fully-connected XX model}
\label{s:canonical}

On the basis of the microcanonical density of states \eqref{e:omegaXX}, the canonical partition function and canonical probability distribution functions of the fully-connected XX model can be computed. The physical motivation for such calculations stems from the possibility to weakly couple the earlier described Bose--Einstein condensate realizations of LMG-type models to large assemblies of ``bath'' atoms, or to decouple at a certain instance of time a small part of the condensate from the rest, such that ``subsystem thermalization'' may account for the subsystem being described by a canonical ensemble. A more detailed discussion of the experimental feasibility is given in Sec.~\ref{s:Conclusions}.

The canonical partition function can be expressed in terms of the microcanonical density of states as
\begin{equation}
Z(\beta)=\int\Omega(\varepsilon)e^{-N\beta \varepsilon}d\varepsilon.
\end{equation}
If, as is usually the case, $\Omega$ grows exponentially with $N$, then the entropy $S$ is extensive and its density $s=S/N$ has a finite large-$N$ limit. The partition function	
\begin{equation}\label{e:Zext}
Z(\beta)=\int e^{N[s(\varepsilon)-\beta \varepsilon]}d\varepsilon
\end{equation}
then takes on the form of a Laplace integral \cite{BenOrs}, a property that is closely linked to a large deviation principle \cite{Touchette09}. In the large-$N$ limit, the integrand of \eqref{e:Zext} becomes increasingly sharply peaked, and the value of the integral is entirely dominated by the maximum of the exponent in the integrand. This maximum, in turn, is determined by an interplay of entropic and energetic effects in the model of interest.

For our model, the situation is different: The entropy scales sublinearly with $N$ \eqref{e:sublinS}, entropy and energy no longer contribute on an equal footing, and in the large-$N$ limit their interplay is dominated by the extensive energy. Making use of Eq.~\eqref{e:sublinS}, the canonical partition function can be written as
\begin{equation}
Z(\beta)=N\int e^{\ln\omega_h(\varepsilon)-N\beta \varepsilon}d\varepsilon.
\end{equation}
The energy $\varepsilon$ at which a balanced interplay of the entropic contribution $\ln\omega_h$ and the energetic contribution $N\beta\varepsilon$ will happen, and where hence most of the interesting statistical physics can be observed, is pushed more and more towards the ground state energy with increasing system size $N$. A way of keeping the two terms balanced is to consider the canonical partition function (and also other canonical quantities) as a function of rescaled temperature $\tilde{\beta}=N\beta$, such that
\begin{equation}\label{e:Ztilde}
Z(\tilde{\beta})=N\int e^{\ln\omega_h(\varepsilon)-\tilde{\beta} \varepsilon}d\varepsilon.
\end{equation}
The integrand in \eqref{e:Ztilde} is now independent of $N$ and admits a well-defined and nontrivial large-$N$ limit. Rescaling temperature may look like a cheap trick, but it is just a way of zooming into the energy range at which an interplay of entropic and energetic terms, and hence the interesting physics, takes place. Based on the thus obtained result in the large-$N$ limit, predictions about very large, but finite, systems can be made by transforming back to ``real'' inverse temperature $\beta$.

For the fully-connected XX model \eqref{e:XXcollective}, a straightforward calculation yields
\begin{equation}\label{e:Z}
Z_h(\tilde\beta)=\frac{N}{2}e^{-\tilde{\beta} B_0}\sqrt{\frac{\pi}{2\tilde{\beta}}}\left[\erf\,\Biggl((1+h)\sqrt{\frac{\tilde{\beta}}{2}}\Biggr)+\erf\,\Biggl((1-h)\sqrt{\frac{\tilde{\beta}}{2}}\Biggr)\right]
\end{equation}
for the canonical partition function, where
$\erf$ denotes the error function \cite{NIST}. Conventionally in statistical mechanics, mean values of physical observables and variances of physical observables, like the average magnetization or the magnetic susceptibility, are calculated by taking derivatives of the partition function. While this can be done without too much effort on the basis of \eqref{e:Z}, the result is not particularly enlightening, and much less useful than in the more familiar setting of an extensive entropy. The reason for that is the fact that the canonical probability distribution, which is the measure on Hilbert space with respect to which ensemble averages of observables are computed, does not become exponentially sharply peaked for large $N$. Hence, unlike in the conventional case, distribution functions are no longer characterized by their mean value and variance. Instead, the full probability distribution, or at least many of its moments, are needed to account for the largely fluctuating outcomes, even in the limit of large system size. This is essentially a consequence of the subextensivity of the entropy combined with the rescaling of temperature, which implies that the integral in \eqref{e:Ztilde} is no longer of Laplace type and no large deviation principle is satisfied.

\begin{figure}\centering
  \includegraphics[width=0.95\linewidth]{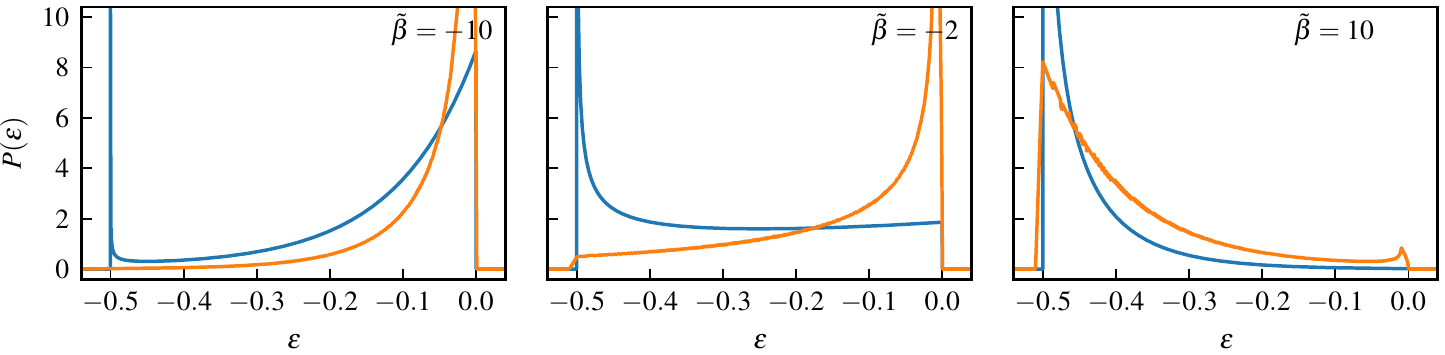}
  \caption{\label{f:pe}
  Canonical probability distributions $P(\varepsilon)$ of the energy density $\varepsilon$ for the fully-connected XX model (blue) and fully-connected TFIM (orange) for zero magnetic field $h$ and inverse temperatures $\tilde\beta$ as indicated in the plots. The results for the fully-connected XX model are obtained on the basis of the density of states \eqref{e:omegaXX} in the large-$N$ limit. The data for the fully-connected TFIM are generated numerically for parameter values $N=5000$ and $\Delta=0.01$. 
  }
\end{figure}

The canonical probability distribution for the outcome density $o$ of an observable $O$ is given by
\begin{equation}\label{e:Po}
P(o)=\Tr\left(\delta(No-O)\frac{\exp(-\beta H)}{Z_h(\beta)}\right),
\end{equation}
where, in the case of the constrained spin models we are considering, $\Tr$ denotes the trace over the constrained Hilbert space. For the canonical probability distribution of the energy density $\varepsilon$ of the fully-connected XX model \eqref{e:XXcollective} we obtain
\begin{equation}\label{eq:Pe}
P(\varepsilon)=N\omega_h(\varepsilon)\frac{\exp(-\tilde{\beta} \varepsilon)}{Z_h(\tilde{\beta})},
\end{equation}
with $\omega_h$ and $Z_h$ as defined in \eqref{e:omegaXX} and \eqref{e:Z}. The distribution of the magnetization density $m_z$ corresponding to the spin component $S^{z}$ is given by
\begin{equation}\label{eq:Pm}
P(m_z)=\begin{cases}
N\exp[-\tilde\beta \varepsilon_{m_z,h}]/Z_h(\tilde\beta) & \text{for $|m_z|\leq 1/2$},\\
0 & \text{otherwise},
\end{cases}
\end{equation}
with
\begin{equation}
\varepsilon_{m_z,h}=2m_z^2-2hm_z-\frac{1}{2}-\frac{1}{N}.
\end{equation}
Examples of these distributions are plotted in Figs.~\ref{f:pe} and \ref{f:pm}. Keeping in mind that all these probability distributions are valid also in the limit $N\to\infty$, it is evident that the distributions do not become arbitrarily sharply peaked and concentration of measure does not happen. This has remarkable and rather uncommon implications for measurements of energy or magnetization in such a system.  Referring, for example, to the distribution $P(\varepsilon)$ for $\tilde\beta=-2$ in the center panel of Fig.~\ref{f:pe}, making measurements on multiple copies of systems in a canonical Gibbs state would result in single-shot energies likely to be spread over the entire range $[-1/2,0]$ of accessible energy densities $\varepsilon$, with the maximum and minimum values even occurring with particularly high probability. Energy densities around the mean value of the distribution, which is located around $\varepsilon=-0.245$, are not especially likely, and are certainly not a particularly good way of characterizing the outcome of a series of energy measurements.

\section{Fully-connected transverse-field Ising model}
\label{s:fcTFIM}

\begin{figure}[b]\centering
		\includegraphics[width=0.95\linewidth]{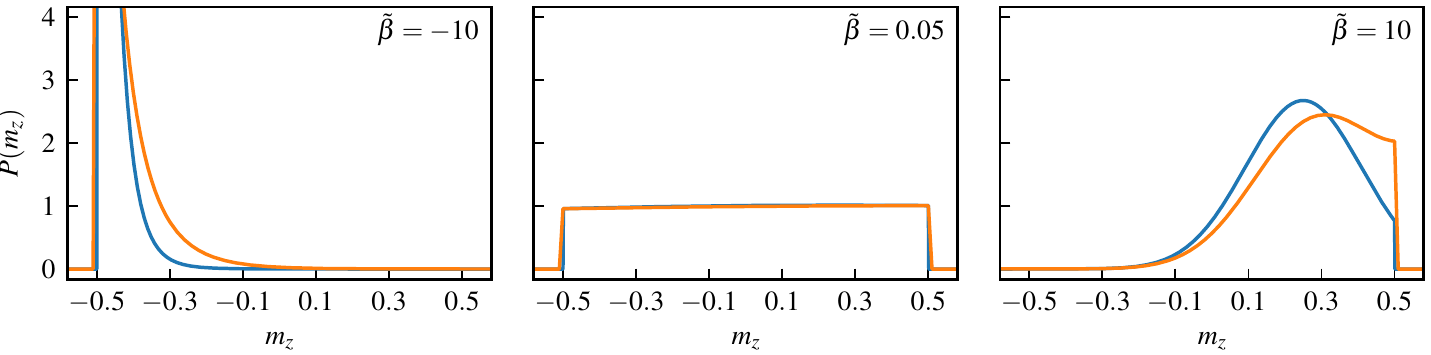}
	\caption{\label{f:pm} 
	Canonical probability distributions $P(m_z)$ of the magnetization density $m_z$ for the fully-connected XX model (blue) and the fully-connected TFIM (orange). The plots are for magnetic field $h=1/2$ and inverse temperatures as indicated in the plots. Data were generated as specified in the caption of Fig.~\ref{f:pe}.
	 The peaks of the distributions in the left panel are of finite height.
	}
\end{figure}

In Secs.~\ref{s:DoS} and \ref{s:canonical} we have discussed the main theoretical concepts of this work on the basis of the fully-connected XX model \eqref{e:XXcollective}, which is particularly amenable to analytic calculations. The experiments with Bose-Einstein condensates that motivated the present study of constrained spin models, however, are described by the fully-connected transverse-field Ising model \eqref{e:Hcollective} or, equivalently, Hamiltonian \eqref{e:LMGcollective} with $\gamma=0$,
\begin{equation}\label{e:TFIMcollective}
H=-\frac{2}{N}(S^{x})^2-2hS^{z}.
\end{equation}
In this case, the angular momentum eigenstates $|N/2,M\rangle$ are no longer eigenstates of the Hamiltonian. With somewhat more sophisticated techniques, it is still possible to derive exact analytic expressions for the density of states of the constrained spin model for arbitrary $\gamma$ in the large-$N$ limit \cite{RibeiroVidalMosseri07}. Alternatively, one can deal with finite system sizes by writing the Hamiltonian in matrix form in the constrained angular momentum eigenbasis $\left\{|N/2,M\rangle\right\}_{M=-N/2}^{N/2}$ (as in Refs.~\cite{CanevaFazioSantoro08,SantosTavoraPerezBernal16}) and numerically diagonalize the resulting matrix of size $(N+1)\times(N+1)$. We opted for this latter approach where, due to the linear increase of the Hilbert space dimension with $N$, the experimentally relevant condensate sizes of hundreds or thousands of bosons are easily accounted for. Plots of the thus calculated densities of states $\omega_h(\varepsilon)$ for the fully-connected TFIM are shown in Fig.~\ref{f:TFIM_dos}. Comparing with the density of states of the fully-connected XX model in Fig.~\ref{f:dosnum}, the main difference is that for the fully-connected TFIM the location of the peaks is not restricted to the boundaries of the energy spectrum. Instead, a peak can occur anywhere in the spectrum, and is moving from high to low energies as $h$ increases (see \cite{RibeiroVidalMosseri07} for a more detailed account). 

\begin{figure}\centering
		\includegraphics[width=0.95\linewidth]{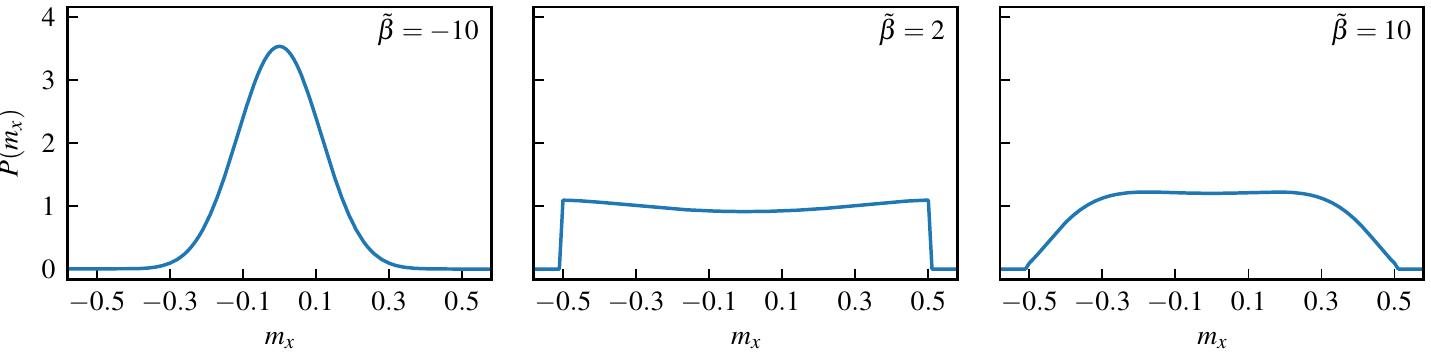}
	\caption{\label{f:lmg_px} Canonical probability distributions $P(m_x)$ of the magnetization density $m_x$ of the fully-connected TFIM for $N=100$. The plots are for magnetic field $h=1/2$ and for rescaled inverse temperatures $\tilde{\beta}$ as indicated in the plots. Data were generated as specified in the caption of Fig.~\ref{f:pe}. The distribution changes from singly peaked (left) to doubly peaked (center and right), somewhat in analogy to the Landau free energy functional in the vicinity of a continuous phase transition.
	}
\end{figure}

The corresponding canonical probability distribution functions, calculated according to \eqref{e:Po}, are shown in Figs.~\ref{f:pe} and \ref{f:pm}. Despite the differences in the densities of states just discussed, the canonical probability distributions $P(\varepsilon)$ and $P(m_z)$ of the fully-connected TFIM (shown in orange) share, at least qualitatively, the relevant features of the XX model (shown in blue), including the doubly peaked structure of $P(\varepsilon)$ at intermediate values of the rescaled inverse temperature $\tilde\beta$. Also, none of the distributions shows concentration of measure in the large-$N$ limit, similarly to what was discussed for the fully-connected XX model at the end of Sec.~\ref{s:canonical}.

\section{Phase transition-like behavior of the magnetization}
\label{s:PT}

Concentration of measure, i.e., the exponential sharpening of distribution functions, can be seen as an essential requirement for a phase transition to take place. Linked to this is the fact that a satisfactory rigorous definition of a phase transition has to make reference to the large-system limit in which distributions are ``infinitely sharp'' and can account for nonanalytic behavior of expectation values of observables. The absence of concentration of measure in our constrained spin models, as discussed in the final paragraphs of Secs.~\ref{s:canonical} and \ref{s:fcTFIM}, therefore precludes the occurrence of a phase transition in the usual, rigorous sense. Since the constrained and the unconstrained models share the same Hamiltonian, it is nonetheless tempting to look for hints or precursors of a behavior that, though not a transition in the rigorous sense, shares at least some of its features. 

\begin{figure}\centering
	\includegraphics[width=0.6\linewidth]{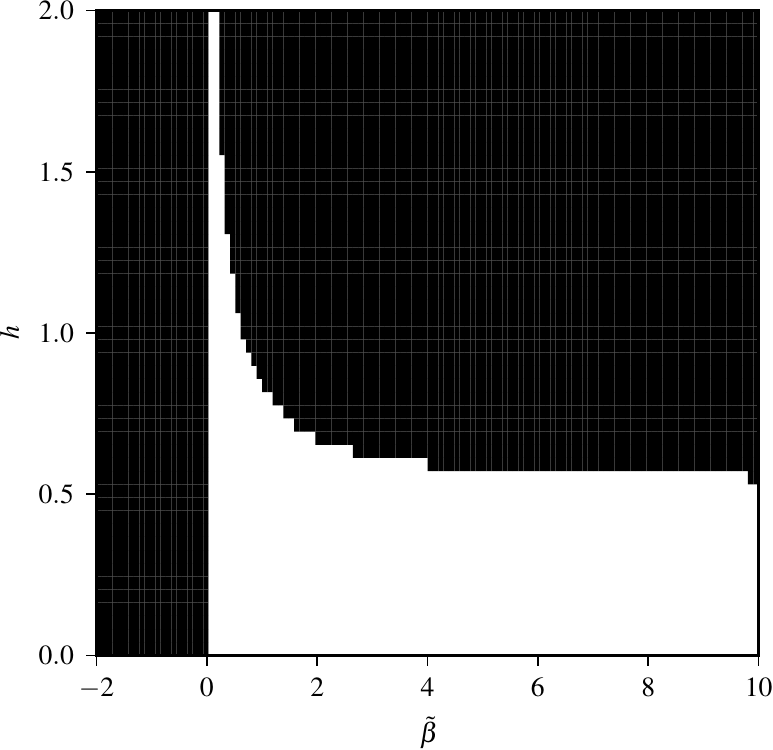}
	\caption{\label{f:lmg_phase} ``Phase diagram'' in the $(T,h)$-plane for the constrained fully-connected TFIM, based on whether the canonical distribution function $P(m_x)$ is singly peaked (black) of doubly peaked (white).}
\end{figure}

The unconstrained fully-connected TFIM, i.e., Hamiltonian \eqref{e:TFIMcollective} without any constraint on the Hilbert space, shows a continuous phase transition at some critical inverse temperature $\beta_\text{c}$ from a paramagnetic phase at $\beta<\beta_\text{c}$ to a ferromagnetic phase at $\beta>\beta_\text{c}$. The order parameter signaling this transition is the magnetization in $x$ direction, which is nonzero in the ferromagnetic phase and vanishes in the paramagnetic phase. To search for similar behavior in the constrained version of the fully-connected TFIM, we show in Fig.~\ref{f:lmg_px} its canonical probability distribution $P(m_x)$ corresponding to the component $S^{x}$ of the collective spin operator. The distribution is singly peaked at small values of the rescaled inverse temperature $\tilde\beta$, becomes flatter, and changes to a doubly peaked distribution for larger $\tilde\beta$. Such a behavior is reminiscent of the shape of the Landau free energy in the vicinity of a continuous phase transition, where the change from a single to a double peak marks the transition point. In the constrained spin models considered in the present work, the absence of concentration of measure implies that the change between singly and doubly peaked distributions has less drastic effects on expectation values of observables, and in particular does not lead to order parameter-like, nonanalytic behavior of the magnetization. Instead, in constrained spin models the observation of such transition-like behavior has to be based on the full probability distribution function (or at least on several of its moments). Based on this criterion, the phase diagram of the fully-connected constrained TFIM in the $(T,h)$-plane is mapped out in Fig.~\ref{f:lmg_phase}. For any given temperature $T$, the canonical probability distribution $P(m_x)$ becomes singly peaked for sufficiently large magnetic field $h$.

\section{Discussion and Conclusions}
\label{s:Conclusions}

In this paper we have studied statistical physical and thermodynamic properties of fully-connected spin models, subject to a constraint on the Hilbert space. The constraint reduces the Hilbert space such that its dimension scales only linearly with the system size $N$. The study is motivated by experiments with two-component Bose--Einstein condensates, which can be mapped onto such constrained spin models by means of a Schwinger mapping. We discussed some of the consequences of linear scaling of the Hilbert space with $N$, in particular the fact that the relevant probability distributions of statistical physics lack the concentration property, i.e., do not become exponentially sharply peaked in the limit of large $N$. As a consequence, expectation values of observables are not sufficient to characterize such systems, but full distribution functions are required instead. Experimental measurements of such distribution functions in the Bose--Einstein condensate realizations of constrained spin models appear feasible, but require multiple experimental runs. We have computed and discussed canonical distribution functions for the fully-connected XX model (which is easily solvable) and the fully-connected TFIM (which is closer to the above mentioned experiments with two-component condensates). Unconventional features of the distribution functions are observed, like the absence of concentration of measure, and transitions from singly peaked to doubly peaked distribution functions are found.

Having discovered and discussed some unconventional statistical physical properties, one may wonder how realistic the assumption of a canonical distribution is for the discussed constrained spin models with linearly growing Hilbert spaces. A semiclassical analysis in the large-$N$ limit shows \cite{SciollaBiroli11}, and numerical calculations for finite systems confirm, that the dynamics of both models considered in this paper evolve periodically in time with a rather short period, and relaxation to a canonical distribution does not occur. The same holds also true for a small subsystem of a larger LMG model, which evolves periodically with typically the same oscillation period as the total system. Since subsystem equilibration of a ``pure'' LMG model does not occur, coupling to an external bath, like a large assembly of bosons or even classical harmonic oscillators, seems to be needed in order for the constrained spin model to reach a canonical equilibrium state. A theoretical study of an LMG model in a bath would typically be done by making use of the framework of open quantum systems. Deriving a master equation for an LMG model in, say, a bosonic bath is certainly possible, at least under a number of approximations like a Born-Markov and a secular approximation. These approximations, and in particular the secular one, may however be deceiving in that they artificially enforce thermalization to a Gibbs state, even in cases where the total system plus bath in reality does not thermalize \cite{GevaRosenmanTannor00}. A thorough theoretical analysis, with a proper understanding of the effect of the approximations made, goes well beyond the scope of the present paper. This is work in progress and will be reported elsewhere in the future.

Finally, we would like to comment on the feasibility of experimentally probing thermal equilibrium properties of the constrained LMG model. As mentioned in the Introduction, the fully-connected TFIM Hamiltonian with the constraint $S=N/2$ has been successfully engineered with \textsuperscript{87}Rb condensates \cite{Ziebold_etal10,Gross_etal10,Strobel_etal14}. These experiments realize the model as an essentially isolated system, evolving unitarily under the Hamiltonian \eqref{e:Hcollective}, and thermalization to a canonical equilibrium state is not expected. A constrained LMG model in an external bath has not yet been experimentally realized, but adding some sort of thermal bath to the unitarily-evolving experimental set-up, while certainly challenging, does not seem out of range for current state-of-the-art experiments. One such possibility could be the creation of a large two-component \textsuperscript{87}Rb condensate, and a subsequent change of the trapping potential into a double-well, such that a condensate with particle number $S$, which acts as the system, is weakly tunnel-coupled to a similar condensate with particle number $S'\gg S$, which acts as the bath. Such a modification of the trapping potential is a well-established experimental technique \cite{GatiOberthaler07} but, to the best of our knowledge, it is an open question whether such a scenario indeed leads to thermalization of the smaller system. Another established scheme for subjecting a condensate to a bath is by spatially superimposing clouds of two different species of atoms such that one cloud acts as a bath for the other. Such a scheme has for example been realized for a mixture of \textsuperscript{87}Rb and \textsuperscript{171}Yb atoms \cite{Vaidya_etal15}, but not yet for the two-component condensate on which the emulation of the constrained LMG Hamiltonian is based. Finally, a photon bath scattering off the atoms is yet another method for implementing an external bath for a system of ultracold atoms, as realized for example in the \textsuperscript{40}K experiments of Ref.~\cite{Lueschen_etal17}. On the theoretical side, all these realizations of baths may look promising, but more work is needed to establish whether any of them indeed leads to thermalization of the immersed LMG system. On the experimental side, while each of the above schemes has been implemented in some kind of ultracold atom experiment, every applications comes with its own challenges and may need specifically tailored solutions.

\begin{acknowledgements}
The authors acknowledge useful discussions with Markus Oberthaler and Hugo Touchette.
J.R.W.\ acknowledges financial support through the internship programme of the National Institute for Theoretical Physics (South Africa).
M.K.\ acknowledges financial support from the National Research Foundation of South Africa via the Competitive Programme for Rated Researchers.
\end{acknowledgements}


\bibliographystyle{spphys}
\bibliography{LRLR}

%
%

\end{document}